\documentclass[prl,twocolumn,superscriptaddress]{revtex4}
\usepackage{graphicx}
\usepackage{amssymb}
\usepackage{amsmath}
\usepackage{bm}
\usepackage{color}

\begin{document}

\title{Chiral modes at exceptional points in exciton-polariton quantum fluids.}

\author{T. Gao}
\affiliation{Nonlinear Physics Centre, Research School of Physics and Engineering, The Australian National University, Canberra ACT 2601, Australia}

\author{G. Li}
\affiliation{School of Physics and Astronomy, University of Southampton, SO17 1BJ, Southampton, United Kingdom}

\author{E. Estrecho}
\affiliation{Nonlinear Physics Centre, Research School of Physics and Engineering, The Australian National University, Canberra ACT 2601, Australia}

\author{T. C. H. Liew}
\affiliation{Division of Physics and Applied Physics, Nanyang Technological University, Singapore}

\author{D. Comber-Todd}
\affiliation{Nonlinear Physics Centre, Research School of Physics and Engineering, The Australian National University, Canberra ACT 2601, Australia}

\author{A. Nalitov}
\affiliation{School of Physics and Astronomy, University of Southampton, SO17 1BJ, Southampton, United Kingdom}

\author{M. Steger}
\affiliation{Department of Physics and Astronomy, University of Pittsburgh, PA 15260, USA}

\author{K.~West}
\affiliation{Department of Electrical Engineering, Princeton University, Princeton, New Jersey 08544, USA}

\author{L.~Pfeiffer}
\affiliation{Department of Electrical Engineering, Princeton University, Princeton, New Jersey 08544, USA}

\author{D. Snoke}
\affiliation{Department of Physics and Astronomy, University of Pittsburgh, PA 15260, USA}

\author{A. V. Kavokin}
\affiliation{School of Physics and Astronomy, University of Southampton, SO17 1BJ, Southampton, United Kingdom}
\affiliation{SPIN-CNR, Viale del Politecnico 1, I-00133 Rome, Italy}

\author{A. G. Truscott}
\affiliation{Laser Physics Centre, Research School of Physics and Engineering, The Australian National University, Canberra ACT 2601, Australia}

\author{E. A. Ostrovskaya}
\affiliation{Nonlinear Physics Centre, Research School of Physics and Engineering, The Australian National University, Canberra ACT 2601, Australia}

\begin{abstract}
We demonstrate generation of chiral modes -- vortex flows with fixed handedness in exciton-polariton quantum fluids. The chiral modes arise in the vicinity of exceptional points (non-Hermitian spectral degeneracies) in an optically-induced resonator for exciton polaritons. In particular, a vortex is generated by driving two dipole modes of the non-Hermitian ring resonator into degeneracy. Transition through the exceptional point in the space of the system's parameters is enabled by precise manipulation of real and imaginary parts of the closed-wall potential forming the resonator. As the system is driven to the vicinity of the exceptional point, we observe the formation of a vortex state with a fixed orbital angular momentum (topological charge). Our method can be extended to generate high-order orbital angular momentum states through coalescence of multiple non-Hermitian spectral degeneracies, which could find application in integrated optoelectronics.
\end{abstract}
\maketitle

{\em Introduction}. Exceptional points in wave resonators of different origin arise when both spectral positions and linewidths of two resonances coincide and the corresponding spatial modes coalesce into one \cite{Berry2004,Heiss2012}. Originally identified as an inherent property of non-Hermitian quantum systems \cite{Bender2007,Moiseyev2011,Bird15}, exceptional points have become a focus of intense research in classical systems with gain and loss \cite{Wierzig2015}, such as optical cavities \cite{An2009}, microwave resonators \cite{Dembowski2001,Dembowski2003}, and plasmonic nanostructures \cite{PlasmonicsEP2016}. The counterintuitive behaviour of a wave system in the vicinity of an exceptional point led to demonstrations of a range of peculiar phenomena, including enhanced loss-assisted lasing \cite{Peng2014,Rotter2014}, unidirectional transmission of signals \cite{Segev2010}, and loss-induced transparency \cite{Guo2009}.

Due to the nontrivial topology of the exceptional point, the two eigenstates coalesce with a phase difference of $\pm\,\pi/2$, which results in a well-defined handedness (chirality) of the surviving eigenstate \cite{Heiss2001}. This remarkable property of the eigenstate at the exceptional point was first experimentally demonstrated in a microwave cavity \cite{Dembowski2003} and, very recently, led to observation of directional lasing in optical micro-resonators \cite{Peng2016,vortex_laser}. So far, the chirality of the unique eigenstate at an exceptional point has not been demonstrated in any quantum system.

In this work, we demonstrate formation of a chiral state at an exceptional point in a macroscopic quantum system of condensed exciton polaritons. Exciton polaritons are hybrid light-matter bosonic quasiparticles arising due to strong coupling between excitons and photons in semiconductor microcavities \cite{Deng_10,CiutiREV13}. Once sufficient density of exciton polaritons is injected by an optical or electrical pump, the transition to quantum degeneracy occurs, whereby typical signatures of a Bose-Einstein condensate emerge \cite{Deng_02,BEC06,BEC07,Deng_10,CiutiREV13,YamamotoREV14}. Radiative decay of polaritons results in the need for a continuous pump to maintain the population. This intrinsic open-dissipative nature of exciton-polariton condensates offers a new platform for study of non-Hermitian quantum physics. Several recent experiments have exploited the non-Hermitian nature of exciton-polariton systems \cite{Savvidis01,weak_lasing,Lieb,Gao2015}. Importantly, the existence of exceptional points and the associated topological Berry phase has been demonstrated in an optically-induced resonator (quantum billiard) for coherent exciton-polariton waves \cite{Gao2015}.

An optically-induced exciton-polariton resonator is a closed-wall potential arising due to injection of high-energy excitonic quasiparticles by an off-resonant optical pump and strong repulsive interaction between the excitonic reservoir and the condensate \cite{Bloch2010}. The size of the resonator is comparable to the de Broglie wavelength of the condensed exciton polaritons, and its geometry is defined by the spatial distribution of the optical pump \cite{Manni2011,Cristofolini13,Askitopoulos13}. Observation of the exceptional points in the spectra of exciton-polariton resonators is enabled by two characteristic features of this system. First, the optically confined exciton polaritons form a multi-mode condensate, i.e. they can occupy several single-particle energy states of the pump-induced effective potential \cite{Tosi12}. Secondly, the pump-induced potential is non-Hermitian, and both real (energy) and imaginary (linewidth) parts of its complex eigenenergies can be precisely controlled by adjusting parameters of the pump \cite{Gao2015}. As a result, two or more eigenstates of the system can be brought to degeneracy.

\begin{figure}[t]
  \centering
  \includegraphics[width=8cm]{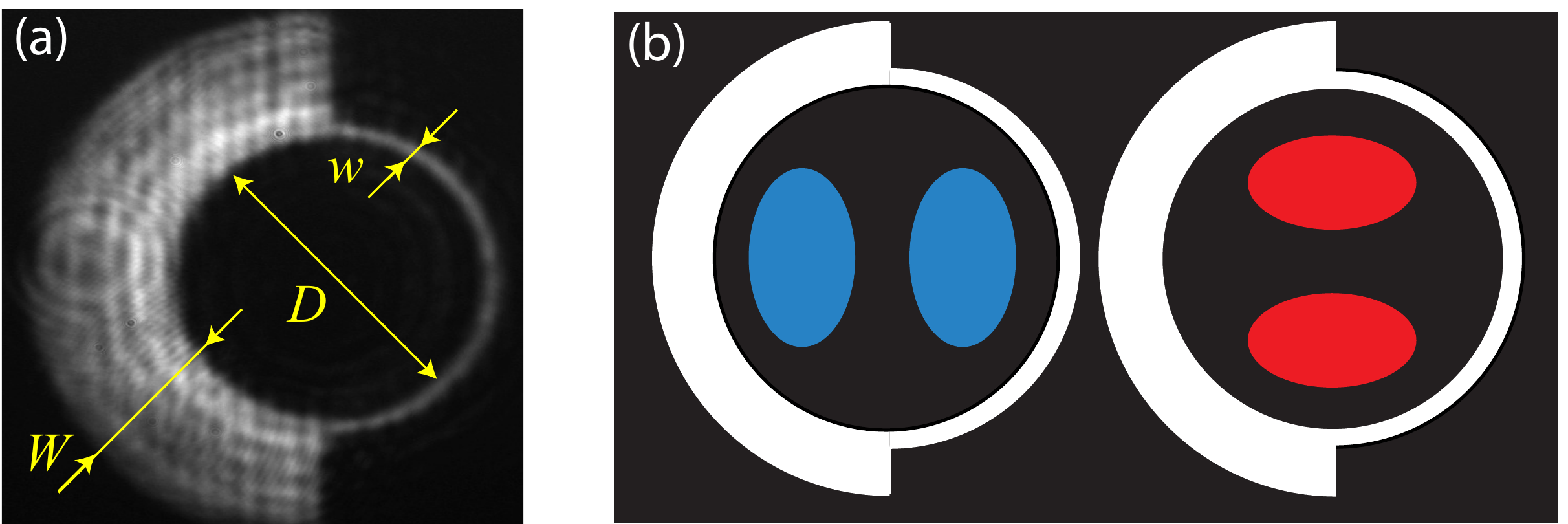}
  \caption{(a) Experimental pump intensity distribution. The width of the left half-ring is $W=9$ $\mu$m and fixed; the width of the right half-ring is tuned between $w_1=1.0$ $\mu$m and $w_2=1.2$ $\mu$m with a step of $0.1$ $\mu$m. The inner ring diameter is $D=15$ $\mu$m and is kept constant. (b) Schematics of the DMD mirror mask shaping the pump spot (white) and the corresponding dipole modes of the trapped exciton polaritons (blue and red).}\label{pump_profile}
\end{figure}

Here, we create a non-Hermitian trapping potential for exciton polaritons in the form of an asymmetric ring resonator, and observe condensation into several trapped modes. By changing the geometry of the pump, and therefore the overlap of the modes with the gain region, we tune the imaginary part of the optically-induced potential and observe the transition from crossing to anti-crossing of complex eigenvalues, which signals the existence of an exceptional point. Furthermore, the high-Q tapered microcavity used for our experiments \cite{SnokePRX2013} enables precise control over the ratio of the exciton and photon in the hybrid quasiparticle. We use this additional control parameter to drive the two lowest-lying dipole states of the system to a vicinity of an exceptional point and confirm the formation of a chiral mode -- a charge one vortex -- in analogy with microwave \cite{Dembowski2003} and optical \cite{Peng2016,vortex_laser} resonators. In addition, we demonstrate formation of a mode with a higher-order topological charge (orbital angular momentum), when two coexisting exceptional points are tuned into close proximity of one another, thus opening the avenue for experimental tests of topological properties of higher-order exceptional points.

\emph{Experiment}. We create exciton polaritons in a high-Q GaAs/AlGaAs microcavity similar to that used in Ref.~\cite{SnokePRX2013}. The details about the experimental setup can be found in {\color{blue} Supplemental Material (SM)}. By utilising a digital micromirror device (DMD), we morph the pump spot into the asymmetric ring shape shown schematically in Fig.~\ref{pump_profile}. The pump simultaneously populates the system with exciton polaritons and forms a closed-wall potential due to the local blue shift in energy induced by the excitonic reservoir \cite{Bloch2010}. Similarly to Ref. \cite{Gao2015}, the inner area of the ring is kept constant and the varying widths of the potential walls, $W$ and $w$, affect the overlap between exciton-polariton condensate modes and the gain region. Analysis of the microcavity photoluminescence by means of energy resolved near-field (real-space) imaging allows us to obtain the spatial density distribution and energy levels corresponding to the condensate modes in the ring resonator. All experiments are performed in the strong coupling regime and the pump power is kept at around $1.5$ times that needed for condensation.

Due to the asymmetry of the potential walls imposed by the pumping geometry, as well as by the cavity gradient \cite{SnokePRX2013}, the eigenmodes of the optically-induced ring resonator resemble the Ince-Guassian modes \cite{Ince_Gaussian}. Once the threshold for the condensation is reached, only a few modes of this resonator are occupied, and we focus on the two lowest-lying dipole modes $(1,1)$ with the orthogonal orientation of the nodal lines. Due to their orientation, the two dipole modes have different overlap with the exciton reservoir and are, in general, not energy degenerate and well separated from the other modes (see {\color{blue} SM}).

\begin{figure}
\centering
\includegraphics[width=8.5 cm]{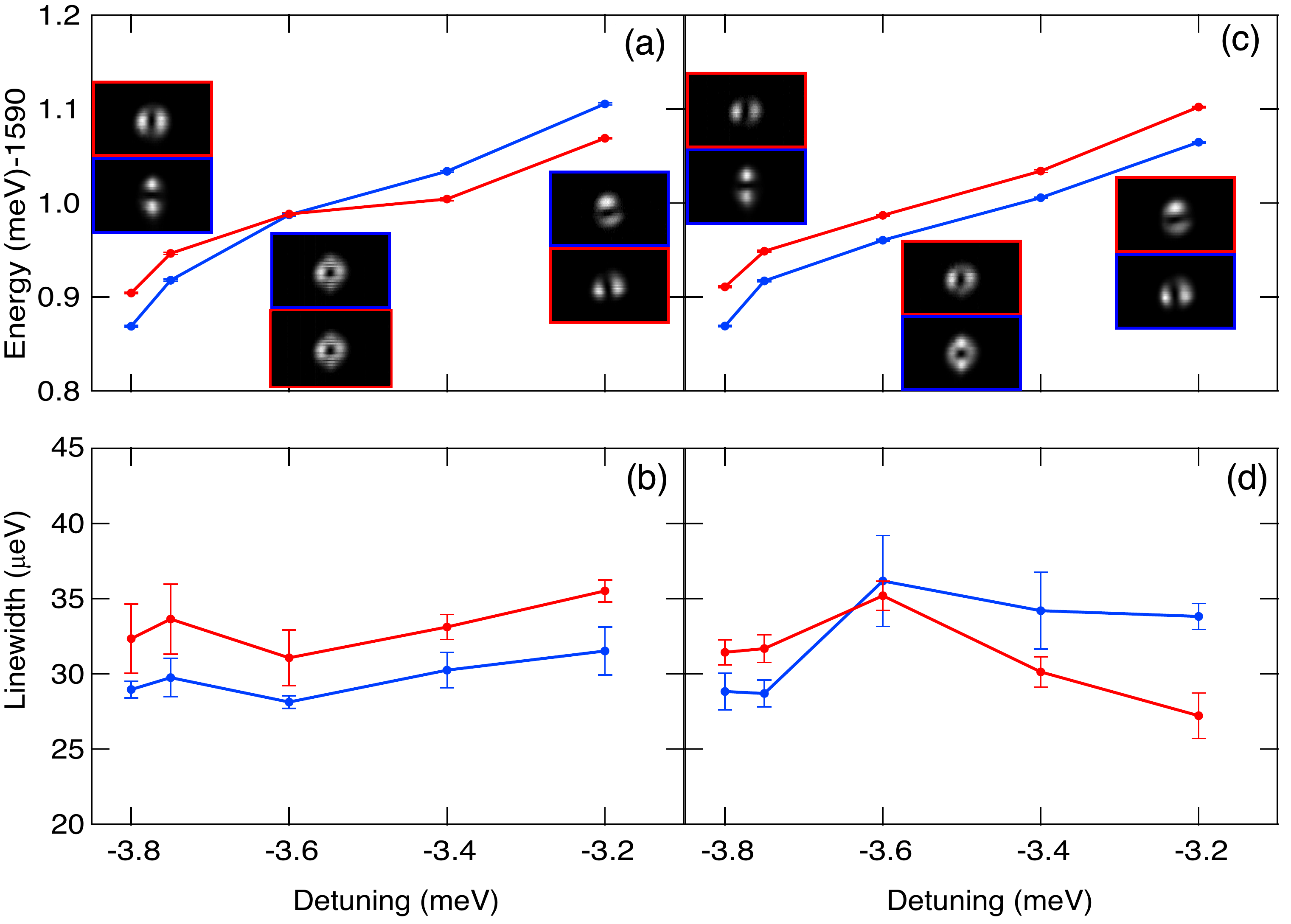}
\caption{ The (a) crossing of the real parts of the eigenergies and (b) anti-crossing of the imaginary parts of the eigenenergies (linewidths) corresponding to the dipole states for the changing detuning, $\Delta$, and fixed width of the right half-ring $w_2=1.2$ $\mu$m. Insets in (a) show the corresponding spatial density structure of eigenmodes observed in the energy-resolved tomography measurement (see {\color{blue} SM}). The (c) anticrossing of the energies and (b) crossing of the linewidths for the changing detuning, $\Delta$, and fixed width of the right half-ring $w_1=1.0$ $\mu$m.} \label{energy_space}
\end{figure}

Because of the non-Hermitian nature of exciton polaritons, the eigenenergy of the modes in the ring potential is also complex, where the real part corresponds to the energy peak position, and the imaginary part corresponds to the linewidth. In our experiment, the peak positions corresponding to the two dipole modes can be tuned by changing the relative admixture of exciton and photon in the exciton polariton. This is achieved by keeping all parameters of the pump fixed and changing the relative position of the excitation beam and the sample, as a sizeable linear variation in the microcavity widths \cite{SnokePRX2013} results in variation of detuning between the cavity photon mode and bare exciton: $\Delta=E_{ph}-E_{ex}$. The change in the energies of the two dipole modes with changing detuning is shown in Fig.~\ref{energy_space}(a), where a clear crossing of the corresponding energy levels can be observed. (The corresponding dispersions below condensate threshold for varying detuning can be seen in {\color{blue} SM}). The transition to spectral degeneracy is accompanied by an avoided crossing of the imaginary parts of the eigenenergy (linewidths), as seen in Fig.~\ref{energy_space}(b). We stress that, due to the high quality factor of the microcavity and long lifetime of exciton polaritons in this microcavity ($\sim~200$ ps), the resonances have a very narrow linewidth, which helps to differentiate between closely positioned energy levels.

\begin{figure}
  \centering
  \includegraphics[width=8cm]{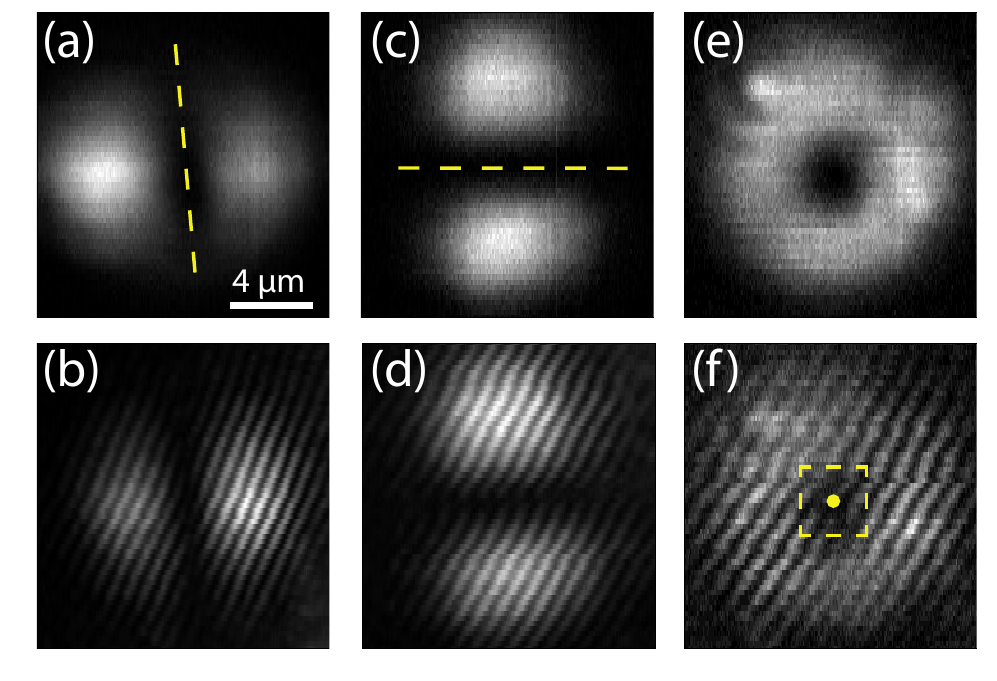}
  \caption{Experimentally observed (a,c,e) probability density distribution of the exciton-polariton condensate and (b,d,f) the corresponding interference pattern for the dipole modes with (a,b) near-vertical and (c,d) near-horizontal nodal line, and (e,f) for a charge one vortex state. The dot in (d) marks the fork in the interference fringes. Parameters are: (a,b,c,d) $\Delta=-3.8$ meV, $w\approx 1.1$ $\mu$m; (e,f) $\Delta=-3.6$ meV, $w\approx1.1$ $\mu$m.
  %{\color{red} Q3: See comment Q2 in the text. Why cannot you do interferometry on the tomography images in Fig. \label{energy_space}(a)?}
  }\label{interference_image}
\end{figure}

From the energy resolved spatial image of the cavity photoluminescence, we reconstruct the spatial probability density distribution of the polariton condensate wavefunction for each detuning value. Away from the degeneracy, the two dipole modes are clearly visible [insets in Fig.~\ref{energy_space}(a)]. As the two complex eigenvalues are tuned in and out of the degeneracy, we observe the characteristic exchange of the modes corresponding to the two energy branches. This kind of mode switch may indicate the existence of an exceptional point, so we vary a second control parameter in our system, which is the width of the right half-ring of the resonator, $w$ (see Fig. \ref{pump_profile}). As the value of this parameter is changed from $w=w_1$ to $w=w_2$, without changing the size of the resonator, $D$, we observe transition from crossing to anticrossing in energy and from anticrossing to crossing in the linewidths of the two resonances corresponding to the dipole modes [see Figs.~\ref{energy_space}(c,d)]. This transition confirms that an exceptional point exists in the parameter space $(\Delta,w)$ \cite{HeissPRE00,Wierzig2015,Gao2015}. Furthermore, by fixing the detuning at $-3.6$ meV, which corresponds to the energy crossing in Fig.~\ref{energy_space}(a), and tuning the right half-ring width from $w_2$ to $w_1$, we observe a vortex-like mode at $w_1<w<w_2$, as shown in Fig.~\ref{interference_image}(e).  

In order to ascertain the nature of this state, we perform interferometry with a magnified ($\times 15$) reference beam derived from a small flat-phase area of the photoluminescence. First of all, the energy-resolved interferometric imaging confirms the phase structure of the two dipole modes Fig.~\ref{interference_image} (a,b) and (c,d) away from the degeneracy point. 
%We note that the lower energy dipole mode has a relatively larger intensity so the total real-space imaging of the modes may imply switching from one dipole mode to the other one. However, the energy- resolved imaging shows that, away from the degeneracy point, there are always two dipole modes. 
Furthermore, the interference pattern shown in Fig.~\ref{interference_image} (f) is stable for many minutes, and reveals a fork in the fringes which is a clear signature of a stable charge-one vortex. This measurement therefore confirms that a vortex with the topological charge one is formed in the vicinity of the spectral degeneracy, which is only possible if the two dipole modes coalesce with the $\pi$/2 phase difference. We note that due to the finite energy (frequency) resolution of spectroscopic measurements in our system, it is impossible to tune the system exactly to the exceptional point, whereby both real and imaginary parts of the complex egenenergy, as well as the eigenstates, would coalesce.

%\emph{Theory}. The physics of exceptional points has been revealed successively in different physical systems, including microwave systems \cite{DembowskiPRL01,DembowskiPRL01}, atomic systems \cite{CartariusPRL07}, optical systems \cite{MiaoScience16}, mechanical systems \cite{DopplerNature16}, and recently polariton systems \cite{Gao 2015}. Following previous works, we will first discuss briefly how to obtain the corresponding $2\times2$ complex Hamiltonian from the dynamical equation, and then discuss the physical mechanism of the generation of multiple vortex states.
{\em Theory.} The full dynamics of an exciton-polariton condensate subject to off-resonant (incoherent) optical pumping can be described by the generalised complex Gross-Pitaevskii equation for the condensate wave function complemented by the rate equation for the density of the excitonic reservoir \cite{Wouters07}:
\begin{equation}\label{GPE}
\begin{aligned}
i \hbar\frac{\partial \psi}{\partial t}=&\left\{-\frac{\hbar^2}{2m}\nabla^2+g_c |\psi|^2+g_R n_R +\frac{i \hbar}{2}\left[ R\, n_R-\gamma_c \right] \right \} \psi, \\
\frac{\partial n_R}{\partial t}=&P(\textbf{r})-\left(\gamma_R+R |\psi|^2\right) n_R,
\end{aligned}
\end{equation}
where $P(\textbf{r})$ is the rate of injection of reservoir particles per unit area and time determined by the pump power and spatial profile of the pump beam, $g_c$ and $g_R$ characterise interaction between condensed polaritons, and between the polaritons and the reservoir, respectively. The decay rates $\gamma_c$ and $\gamma_R$ quantify the finite lifetime of condensed polaritons and the excitonic reservoir, respectively. The stimulated scattering rate, $R$, characterises growth of the condensate density. %The energy relaxation mechanism has not been included in Eq.~(\ref{GPE}) for simplicity, whose effects will be discussed later.

Assuming that, under {\em cw} pumping, the reservoir reaches a steady state,  $n_R(\textbf{r})=P(\textbf{r})/\left(\gamma_R+R |\psi(\textbf{r})|^2\right)$, and that the exciton-polariton density is small near the condensation threshold, Eq.~(\ref{GPE}) transforms into a linear Schr\"odinger equation for the condensate wavefunction, $\psi$, confined to an
effective non-Hermitian potential \cite{Gao2015}:
\begin{equation}\label{effective_potential}
\begin{split}
   V(\textbf{r})=& V_R(\textbf{r})+i V_I (\textbf{r}) \\
                \approx &  -\frac{g_R P(\textbf{r})}{\gamma_R}+i\frac{\hbar}{2}\left[ \frac{R P(\textbf{r})}{\gamma_R}-\gamma_c \right].
\end{split}
\end{equation}
Both real and imaginary parts of this potential depend on the spatial profile of the pump $P(\textbf{r})$, as well as on the exciton-photon detuning, $\Delta$. As discussed in {\color{blue} SM}, the dependence on the detuning enters the model Eq.~(\ref{GPE}) through the dependence of its parameters on the Hopfield coefficient that characterises the excitonic fraction of the exciton polariton \cite{Deng_10}: $|X|^2 =(1/2)\left[ 1+ \Delta/\sqrt{\Delta^2+E_R^2} \right]$, where $E_R$ is the Rabi splitting at zero detuning. 

% the effective mass $m \approx m_{ph}(1-|X|^2)$, the interaction strengths $g_R=|X|^2 g_{ex}$, $g_c=|X|^4 g_{ex}$, and the decay rates $\gamma_c \approx (1-|X|^2)\gamma_{ph}$. Here $m_{ph}\ll m_{ex}$ is the effective mass of the cavity photon, $g_{ex} = 6E_{ex}a_B^2$ characterises interactions between quantum well excitons with the binding energy $E_{ex}$ and the Bohr radius $a_B$ \cite{TassonePRB99}, $\gamma_{ph}$ is the decay rate of the cavity photon, and we assume that the stimulated scattering into the condensed state is more efficient for the more excitonic quasiparticles $R=|X|^2 R_0$. The dependencies of these parameters on detuning for the physically relevant parameters of our system are shown in {\color{blue} SM}.

\begin{figure}
  \centering
  \includegraphics[width=9cm]{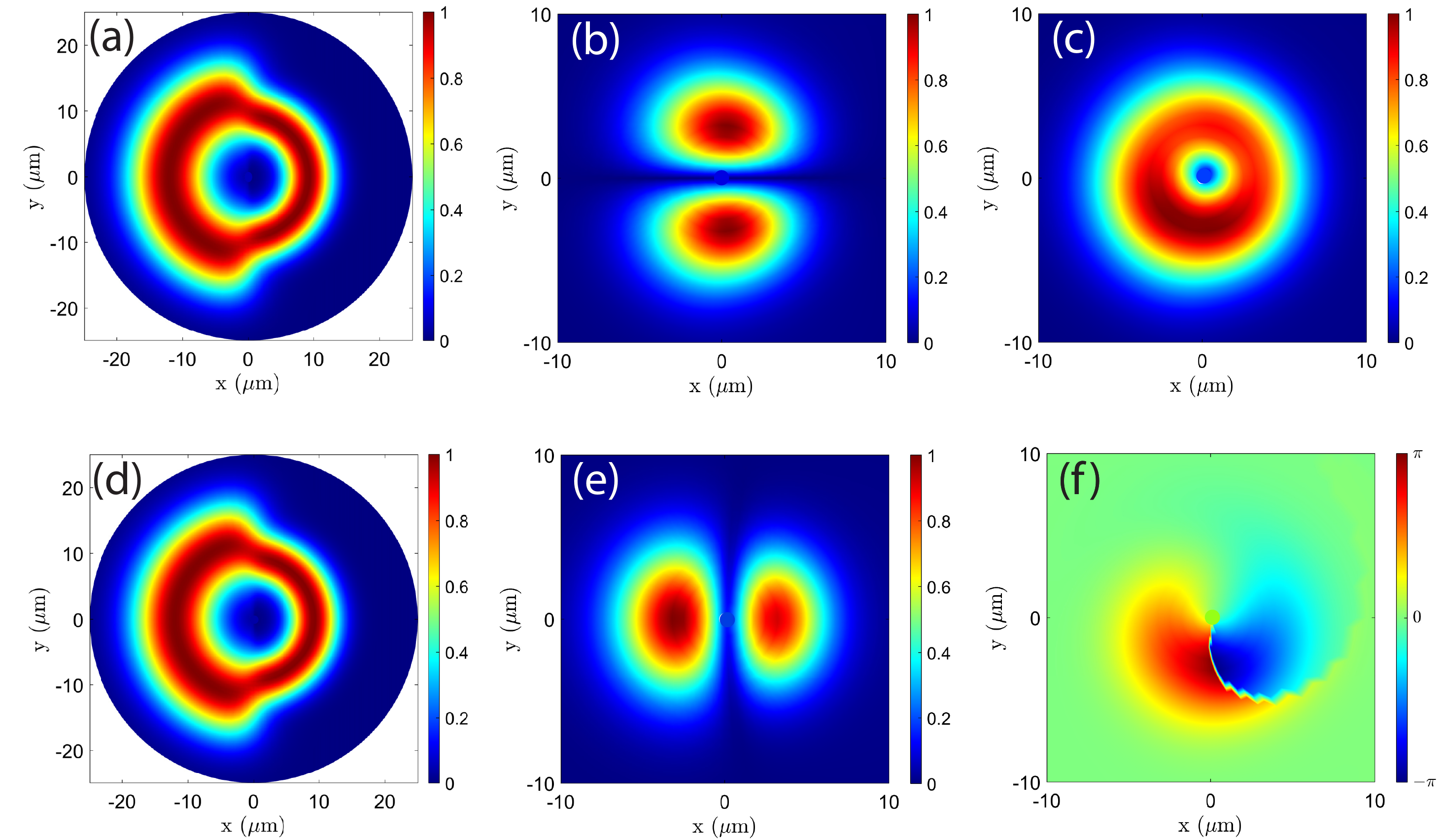}
  \caption{Approximate complex effective potential and the typical profiles of the dipole modes: (a) $V'(\textbf{r})$,  (d) $V''(\textbf{r})$, (b) $|\psi_2|$, (c) $|\psi_2+i\, \psi_3|$, (e) $|\psi_3|$. Panel (f) depicts the phase distribution corresponding to (c).
  %{\color{red} Q5: 1). For what $\Delta$ and $w$ are these calculated? }
  }
\label{Linear_theory_pump_and_modes}
\end{figure}

The eigenstates and complex eigenvalues of the non-Hermitian potential (\ref{effective_potential}) can be found by solving the non-dimensionalised stationary equation:
\begin{equation}\label{Schrodinger equation}
  \left[-\nabla^2 + (V'+iV'') \right]\psi=\tilde{E}\psi.
\end{equation}
where we have introduced the scaling units of length, $L=D$, energy $E_0=\hbar^2/(2m_{ph}L^2)$, and time $T=\hbar/E_0$, $m_{ph}$ is the effective mass of the cavity photon, and $\tilde{E}$ is the complex eigenenergy normalised by $E_0/(1-|X|^2)$. The normalised real and imaginary parts of the potential are: $V'({\bf r})=V'_0 \Gamma_X P({\bf r})/P_0$, $V''({\bf r})=V''_0 \left[\Gamma_X P({\bf r})/P_0-1\right]$, where $V'_0=g_{ex}\gamma_{ph}/(E_0R_0)$, $V''_0=\hbar\gamma_{ph}/(2E_0)$, and $P_0=\gamma_R\gamma_{ph}/R_0$, where $\gamma_{ph}$ is the lifetime of the cavity photon, $g_{ex}$ is the stregth of the exiton-exiton interaction \cite{TassonePRB99}, and we have introduced a base value for the stimulated scattering rate $R_0$ (see {\color{blue} SM}). The dependence on detuning via the Hopfield coefficient is captured by the parameter $\Gamma_X=|X|^2/(1-|X|^2)$. For simplicity, we approximate the shape of $P(\textbf{r})$ by Gaussian envelope functions, with the resulting shapes of $V'({\bf r})$ and $V''({\bf r})$ shown in Fig.~\ref{Linear_theory_pump_and_modes}(a,d).

By solving the eigenequation (\ref{Schrodinger equation}) numerically and sorting values of $\tilde{E}$ in the ascending order of its real part, $E_n$, we obtain the corresponding hierarchy of eigenstates $\psi_n$. Figures~\ref{Linear_theory_pump_and_modes}(b) and (e) show
the moduli of $\psi_2$ and $\psi_3$, respectively, which correspond to the slightly deformed dipole states $(1,1)$ with the orthogonal orientation of nodal lines. Importantly, the existence of these steady states of the exciton polariton condensate in the pump-induced potential is also confirmed by full dynamical simulations of the model equation (\ref{GPE}) (see {\color{blue} SM}).

\begin{figure}
  \centering
  \includegraphics[width=9cm]{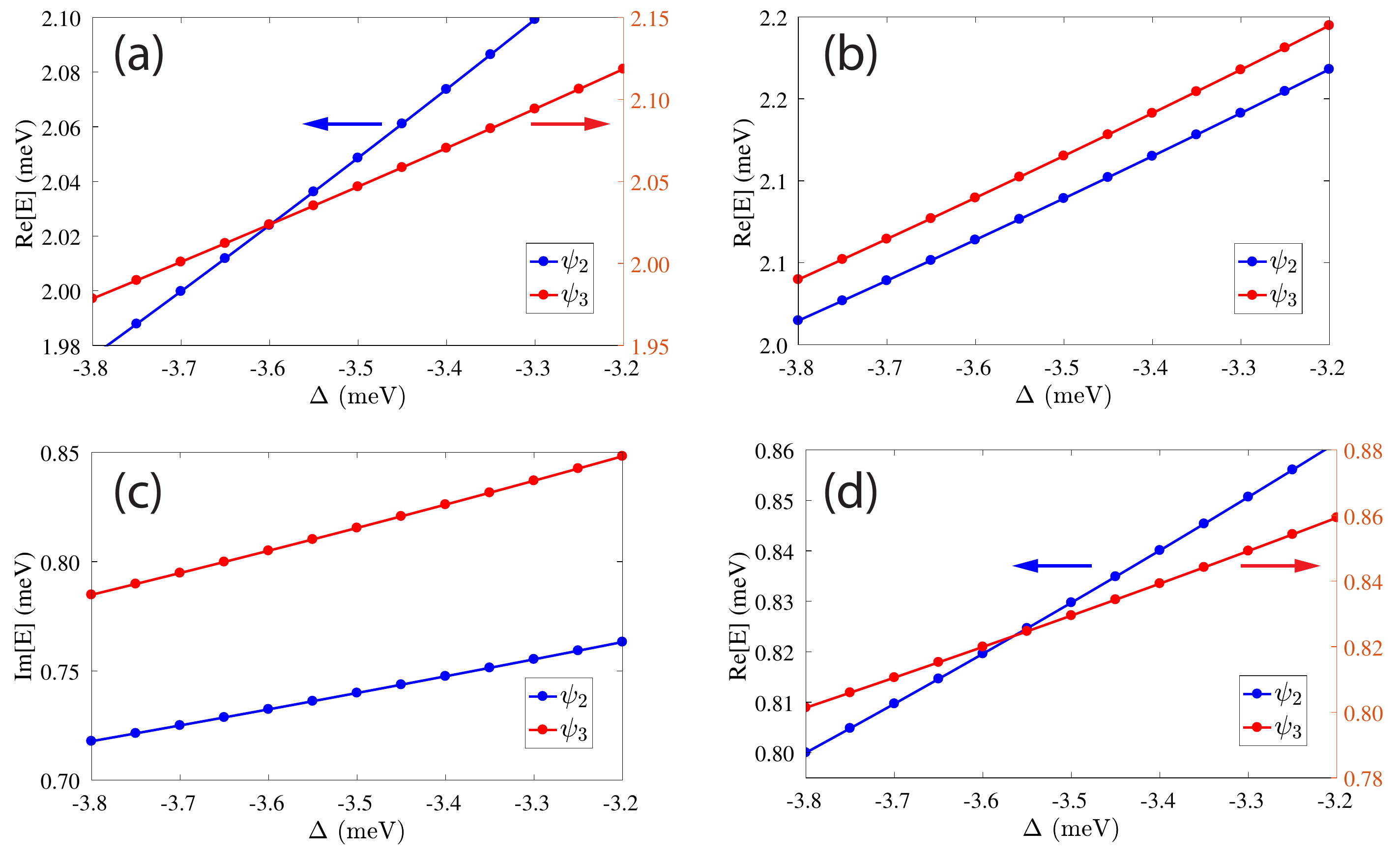}
  \caption{Dependence of energy levels corresponding to the dipole modes on the change of the exciton-photon detuning, $\Delta$. (a)(c) Re$[\tilde{E}]$ crossing and Im$[\tilde{E}]$ anti-crossing at $w=w_2$.
  (b)(d) Re$[\tilde{E}]$ anti-crossing and Im$[\tilde{E}]$ crossing at $w=w_1<w_2$. Panels (a,d) have two different $y$-axes to accentuate the crossing. %The same-axis plots and parameters used in the calculation can be found in {\color{blue} SM}. Arrows indicate the corresponding $y$ axis of the curves.
  %{\color{red} Q6: These should be re-plotted as a function of $\Gamma_X$, or maybe even $\Delta$, since the dependance $\Gamma_X(\Delta)$ is nonlinear. Choose the axis scale so that the crossing is well resolved; currently the curves  are nearly overlapping, which is not acceptable. Give the values of parameters used in these plots.}
  }\label{Linear_theory_cross_anticross}
\end{figure}

The dependence of the real and imaginary parts of the dipole modes eigenenergies on the experimental control parameters, $\Delta$ and $w$, is shown in Fig.~\ref{Linear_theory_cross_anticross}. We can see that our simple linear model reflects the qualitative behaviour observed in the experiment (Fig.~\ref{energy_space}). 
%Note that, for the sake of clarity, Fig.~\ref{Linear_theory_cross_anticross} is plotted for the set of parameters which exaggerates the crossing and the anticrossing behavior. %Otherwise, because of being close to an EP those curves will overlap with each other which will obscure the result.

To understand why the experimental control parameters $\Delta$ and $w$ allow us to tune the system in and out of the vicinity of the exceptional point, we follow the standard approach \cite{BliokhPRL08,Heiss2012,Wierzig2015,Gao2015}, and construct a phenomenological couple-mode model for the two modes with the quantum numbers $n$ and $n'$ near degeneracy (see {\color{blue} SM} for the details). The resulting effective two-mode interaction Hamiltonian can be written as follows:
\begin{equation}\label{two_level_hamiltonian}
  \hat{H}=\left[
                   \begin{array}{cc}
                     \tilde{E}_n & q \\
                     q^* & \tilde{E}_{n^\prime} \\
                   \end{array}
                  \right], \quad\quad \tilde{E}_{n,n^\prime}=E_{n,n^\prime}-i\Gamma_{n,n^\prime},
\end{equation}
where $\tilde{E}_{n,n^\prime}$ are the complex eigenenergies of the uncoupled modes and $q$ characterises their coupling strength.
The eigenvalues of the Hamiltonian (\ref{two_level_hamiltonian}) are $\lambda_{n,n'}=\tilde{E}\pm\sqrt{\delta\tilde{E}^2+|q|^2}$, where $\tilde{E}=(\tilde{E}_n+\tilde{E}_{n'})/2\equiv E+i\Gamma$, and $\delta\tilde{E}=(\tilde{E}_n-\tilde{E}_{n'})/2\equiv \delta E-i\delta\Gamma$. The real and imaginary parts of $\lambda_{n,n'}$ form Riemann surfaces with a branch-point singularity in the space of parameters $(\delta E,\delta\Gamma)$ \cite{Gao2015}, as shown in {\color{blue} SM}. At the exceptional points, $i\delta\tilde{E}_{EP}=\pm|q|$, the eigenvalues coalesce, $\lambda_n=\lambda_{n'}$. The eigenstates also coalesce and form a single chiral state \cite{Heiss2001,Heiss2012}. In our system, the two eigenstates corresponding to $n=2$ and $n'=3$ are dipole modes, and therefore the chiral state is a vortex with a well-defined topological charge one, as shown in Fig.~\ref{Linear_theory_pump_and_modes}(c,f) and Fig.~\ref{interference_image}(c,d).
%
%%The two experimental parameters of our system that control the approach to an exceptional point are the exciton-photon detuning, $\Delta$, and the width of the pump-induced reservoir, $w$. 
%

The two parameters $(\delta E,\delta\Gamma)$ can be related to the experimental parameters $(\Delta,w)$. As discussed in {\color{blue} SM}, increasing $\Delta$ corresponds to increasing $\delta E$ and moves the system away from the spectral degeneracy, while growing $w$ corresponds to decreasing $\delta\Gamma$. Therefore the variable detuning and the width of the resonator's wall allow us to control the approach to the exceptional point as demonstrated in the experiment (Fig. \ref{energy_space}) and confirmed by theory (Fig. \ref{Linear_theory_cross_anticross}).

{\em Conclusion.} In summary, we have experimentally demonstrated the chirality of an eigenstate of a non-Hermitian macroscopic quantum coherent system of Bose-condensed exciton polaritons in the vicinity of an exceptional point. In our experiment, the chiral eigenstate is a vortex with a well-defined, deterministic topological charge (orbital angular momentum). We stress that, contrary to the previous demonstration of an exciton polariton vortex with a well-controlled charge \cite{chiral_lens}, the shape of the optically induced ring resonator that creates a non-Hermitian potential for exciton polaritons in this work does not break the chiral symmetry. In such case, one would expect that the sign of the vortex charge would change randomly between realisations of the experiment \cite{Yulin,Deveaud}, which is not the case here. The deterministic nature of the vortex charge has been confirmed in our experiment by blocking the pump for a sufficient time interval (up to $1$ hour) to let the pump-injected reservoir disappear. After the pumping is resumed, the charge of the vortex remains the same. The controlled and reliable generation of a chiral state with a prescribed topological charge could find use in polariton-based integrated optoelectronic devices.

%three vortex
%\begin{figure}
%  \centering
%  \includegraphics[width=8cm]{./Three_vortex}
%  \caption{(a) Experimental density of the three-vortex state. (b) Experimental interference image of the three-vortex state. (c)(d) Theoretically calculated three-vortex state $\psi=\psi_7+i\,\psi_8+\psi_9+i\,\psi_{10}$, showing the (c) modulus and (d) phase of $\psi$. As a proof of concept, eigenmodes $\psi_n$ were obtained from a ring trap.
%  %{\color{red} Q7: in all plots of the phase in this paper, it is better to use a different colour map, e.g. 'jet'. This new Matlab colourmap is really terrible for representing the phase!}
%  }\label{three_vortex_fig}
%\end{figure}

%two vortex
\begin{figure}
  \centering
  \includegraphics[width=8.5cm]{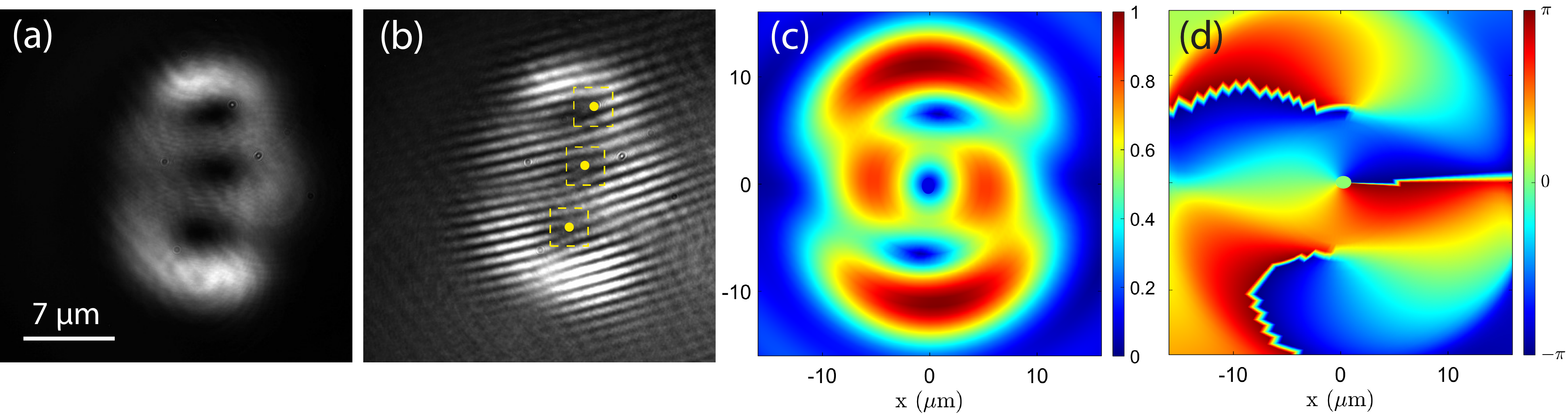}
  \caption{Experimentally measured (a) density and (b) interference image of the triple-vortex state. Theoretically calculated (c) modulus and (b) phase of the triple-vortex state $\psi=\psi_7+i\,\psi_8+\psi_9+i\,\psi_{10}$. The three dots in (b) mark the forks in the interference fringes.
  %{\color{red} Q7: in all plots of the phase in this paper, it is better to use a different colour map, e.g. 'jet'. This new Matlab colourmap is really terrible for representing the phase!}
  }\label{three_vortex_fig}
\end{figure}

Last but not least, our technique of generating chiral states can be applied to higher-order modes in the non-Hermitian exciton-polariton system. In particular, if the inner size of the ring, $D$, is increased, higher-order modes can be populated by the condensate and brought to degeneracy by varying control parameters of the system. Moreover, two (or more) exceptional points can be simultaneously created and brought to a close vicinity of each other. As an example, the higher-order orbital momentum state formed in our experiment by hybridisation of two chiral modes, $\psi=\psi_7+i\,\psi_8+\psi_9+i\,\psi_{10}$, is shown in Fig. \ref{three_vortex_fig}, together with the prediction of our linear theory. The interferometry image confirms that the vortices in the triple-vortex state have the same topological charge (see also a double-vortex state shown in {\color{blue} SM}).
These results offer further opportunities to explore the physics of higher-order exceptional points \cite{EP3_1,EP3_2,EP3_3,EP3_4} and exceptional points clustering \cite{EP3_5,EP_clustering} in an open quantum system.

%Moreover, two (or more) exceptional points can be simultaneously created in the system by tuning the control parameters, and these can be brought to a close vicinity of each other. The higher-order orbital momentum state formed in our experiment by hybridisation of two chiral modes, $\psi=\psi_7+i\,\psi_8+\psi_9+i\,\psi_{10}$, is  shown in Fig. \ref{three_vortex_fig}, together with the prediction of our linear theory. The interferometry image confirms that the three resulting vortices have the same topological charge (see also a two-vortex state shown in {\color{blue} Supplemental Material}). These results offer an exciting opportunity to explore the physics of higher order exceptional points \cite{EP3_1,EP3_2} and exceptional points clustering \cite{EP_clustering} in an open quantum system.

%Fig.~\ref{Linear_theory_cross_anticross} shows the dependence of energy levels $E_2$ and $E_3$ on the effective mass.
%Both crossing and anti-crossing behavior observed in experiments can be reproduced qualitatively. Note that neither in
%Fig.~\ref{Linear_theory_cross_anticross}(a) nor (c) where the crossing occurs corresponding to the vortex mode.
%Because in these cases modes $u_2$ and $u_3$ do not have a fixed $\pi$ relative phase. At the exceptional point two deformed dipole modes maintain a fixed $\pm\pi$ phase and their superposition results in a deformed
%vortex state (Fig.~\ref{Linear_theory_vortex}).

\begin{acknowledgments}

\end{acknowledgments}

%BiBtex reference style
%\bibliographystyle{vortex_gen_ref_format}
%\bibliography{vortex_gen_reference}

%plain tex reference style

\end{document}